\documentclass[final,1p,times]{elsarticle}

\usepackage{graphics}
\usepackage{graphicx}
\usepackage{epsfig}

\usepackage{amssymb}

\usepackage[rightcaption]{sidecap}

\usepackage{subfigure}

\journal{Nuclear Physics A}

\newcommand{\proton}{\ensuremath{\mathrm{p}}{}}
\newcommand{\pip}{\ensuremath{\pi^{+}}{}}
\newcommand{\pim}{\ensuremath{\pi^{-}}{}}
\newcommand{\kp}{\ensuremath{K^{+}}{}}

\newcommand{\pizero}{\ensuremath{\pi^{0}}{}}
\newcommand{\SigmaPlus}{\ensuremath{\Sigma^{+}}{}}
\newcommand{\SigmaZero}{\ensuremath{\Sigma^{0}}{}}
\newcommand{\SigmaMinus}{\ensuremath{\Sigma^{-}}{}}

\newcommand{\LambdaOne}{\ensuremath{\Lambda(1405)}{}} 
\renewcommand{\d}{\ensuremath{\mathrm{d}}}
\newcommand{\dsigmadt}{\ensuremath{\frac{\d \sigma}{\d t}}{}}

\newcommand{\MMsq}{\ensuremath{\mathrm{MM}^{2}}{}}

\begin{document}

\begin{frontmatter}



\title{Properties of the \LambdaOne{} Measured at CLAS}


\author{Kei Moriya and Reinhard Schumacher \\
  for the CLAS collaboration}

\address{Department of Physics, Carnegie Mellon University,
  Pittsburgh, PA 15213, USA}

\begin{abstract}
  Using the CLAS detector system in Hall B at Jefferson Lab, we have
  done a high statistics measurement of the photoproduction of \kp
  $\Lambda(1405)$ using a proton target. The 
  reconstructed invariant mass seen in the various $\Sigma \pi$ decay
  modes has been measured, as well as the
  differential cross section \dsigmadt. The nature of the \LambdaOne{}
  with its peculiar mass distribution is not well understood, and
  studies have shown
  that it may posess strong dynamical components, which could be
  extracted from our data. Various aspects of our analysis are
  discussed, as well as future prospects to further enhance the
  precision of the results.
\end{abstract}


\end{frontmatter}




\section{Introduction}
\label{sec:introduction}

  The \LambdaOne{} is a well-established hyperon resonance located just
  below $N \overline{K}$ threshold. Even though its existence has been
  known since the 1960's and it holds a four-star rating in the
  PDG~\cite{PDG}, its nature as a resonance is not well understood. The
  \LambdaOne{} can only be seen by reconstructing the invariant mass of
  the $\Sigma \pi$ system into which it decays. Previous
  experiments~\cite{Thomas,Hemingway} have shown this mass
  distribution, or ``lineshape''
  to be distorted from the usual Breit-Wigner resonance form, leading to
  various speculations on the nature of the $\Lambda(1405)$.

  Using the CLAS~\cite{CLAS-NIM} detector system in Hall B at the
  Thomas Jefferson National Accelerator Laboratory, we have done a
  photoproduction measurement of the $\Lambda(1405)$. Due to the high
  statistics of the experiment and good resolution of the CLAS system,
  we were able to extract the lineshapes and differential cross
  sections of the \LambdaOne{} in all three of its $\Sigma \pi$ decay
  modes. The lineshapes and cross sections will give a much improved
  quantitative description of the $\Lambda(1405)$, and will lead to further
  insights into its nature and production mechanism.

\section{Theory of the \LambdaOne{} Mass Distribution}
\label{sec:theory-lambdaone-lineshape}

\begin{SCfigure}[][t]
  \centering
  \includegraphics[width=.40\textwidth]{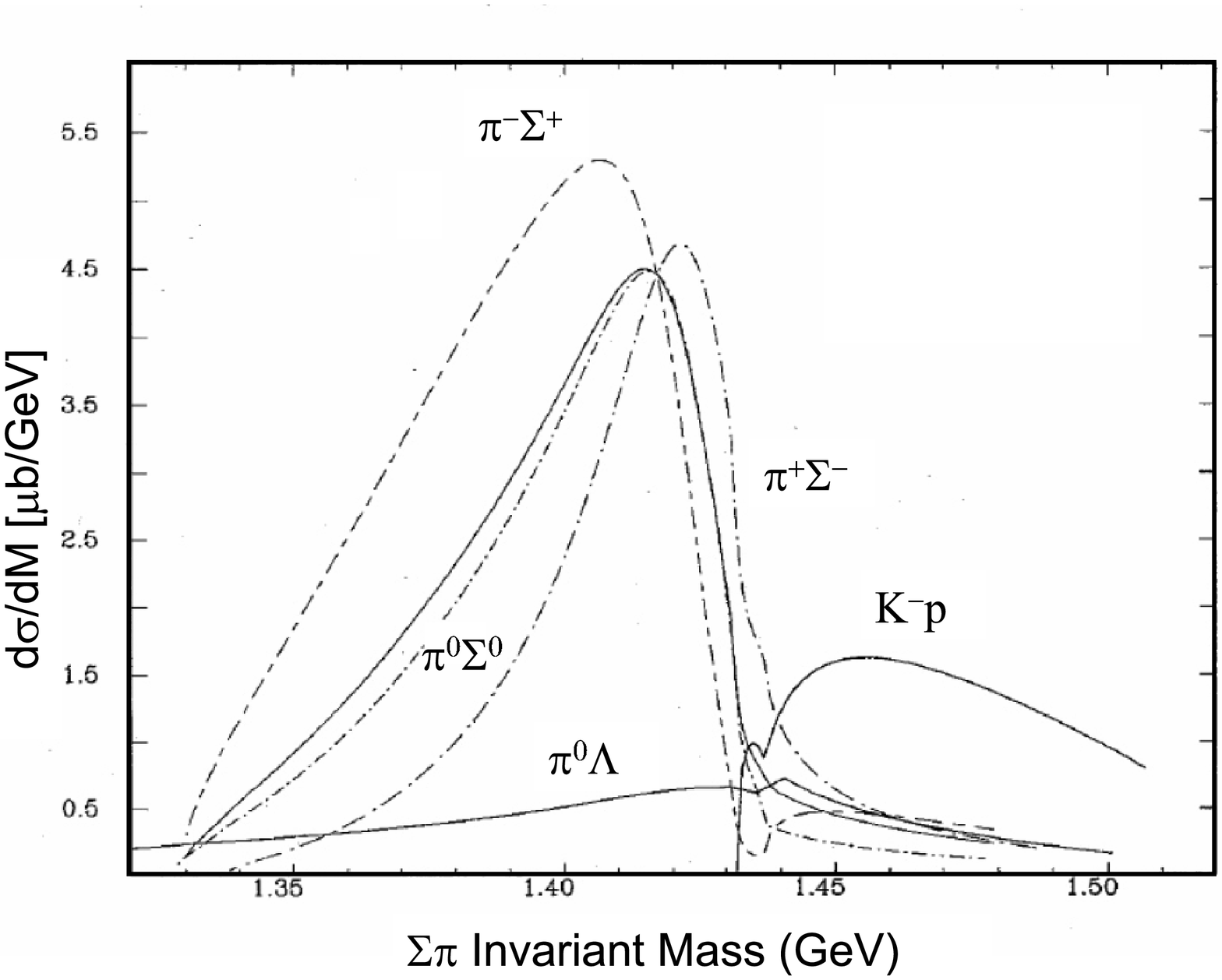}
  \caption{Theoretical lineshapes predicted~\cite{Nacher} for the \LambdaOne{} using
    the chiral unitary approach. Due to the interfering isospin $0$
    and $1$ terms, the lineshapes are distorted from a simple
    Breit-Wigner form. In more recent years, attention has also
    focused on the possible two pole structure of the \LambdaOne{} in
  the $I=0$ channel alone. This too is expected to influence the
  lineshapes~\cite{Jido}.}
  \label{fig:theorylineshape}
\end{SCfigure}

In recent years, theoretical progress in chiral dynamics in the $S=-1$
sector has gained much attention, and has been extended to include a
description of the $\Lambda(1405)$. In this approach the
\LambdaOne{} is dynamically generated by iterating over the meson-baryon
interactions among the ground state baryon octet and psuedoscalar
mesons octet. A paper by Nacher {\it et al.}~\cite{Nacher}
predicts, using a chiral unitary approach, the
lineshapes of the \LambdaOne{} as it decays to its three $\Sigma \pi$
decay modes. Figure~\ref{fig:theorylineshape} shows the predicted
lineshapes, a result of the isospin $0$ and $1$ amplitudes interfering
to produce shifts in the \SigmaPlus \pim{} and \SigmaMinus \pip{} 
lineshapes in opposite directions. Unlike an ordinary resonance where
the lineshape will follow a simple Breit-Wigner form, the \LambdaOne{}
spectrum is predicted to be strongly distorted due to a mixture of
contributions. This figure has been the starting point in our
analysis, and our work has been focused on comparing our experimental
data with this prediction.

\section{Data Analysis Using CLAS}
\label{sec:data-analysis-using-clas}


In this section we show some of the work that was done at CLAS to
study the photoproduction of the $\Lambda(1405)$. The reaction of interest
at CLAS is $\gamma + \proton \rightarrow \kp + 
\Lambda(1405)$, with the
\LambdaOne{} decaying exclusively into $\Sigma \pi$. The CLAS data set
was obtained in 2004 using tagged real photons up to $3.84$ GeV, and a
two charged track trigger.
For the
\SigmaPlus \pim{} decay mode, there are two possible decays for the
\SigmaPlus, i.e. to \proton \pizero or n \pip. The first step of the
analysis is to select either $(\proton,\kp,\pim)$ or $(\kp,\pip,\pim)$
events.

\begin{figure}[htp]
  \centering
  \subfigure[]{
    \includegraphics[width=.40\textwidth]{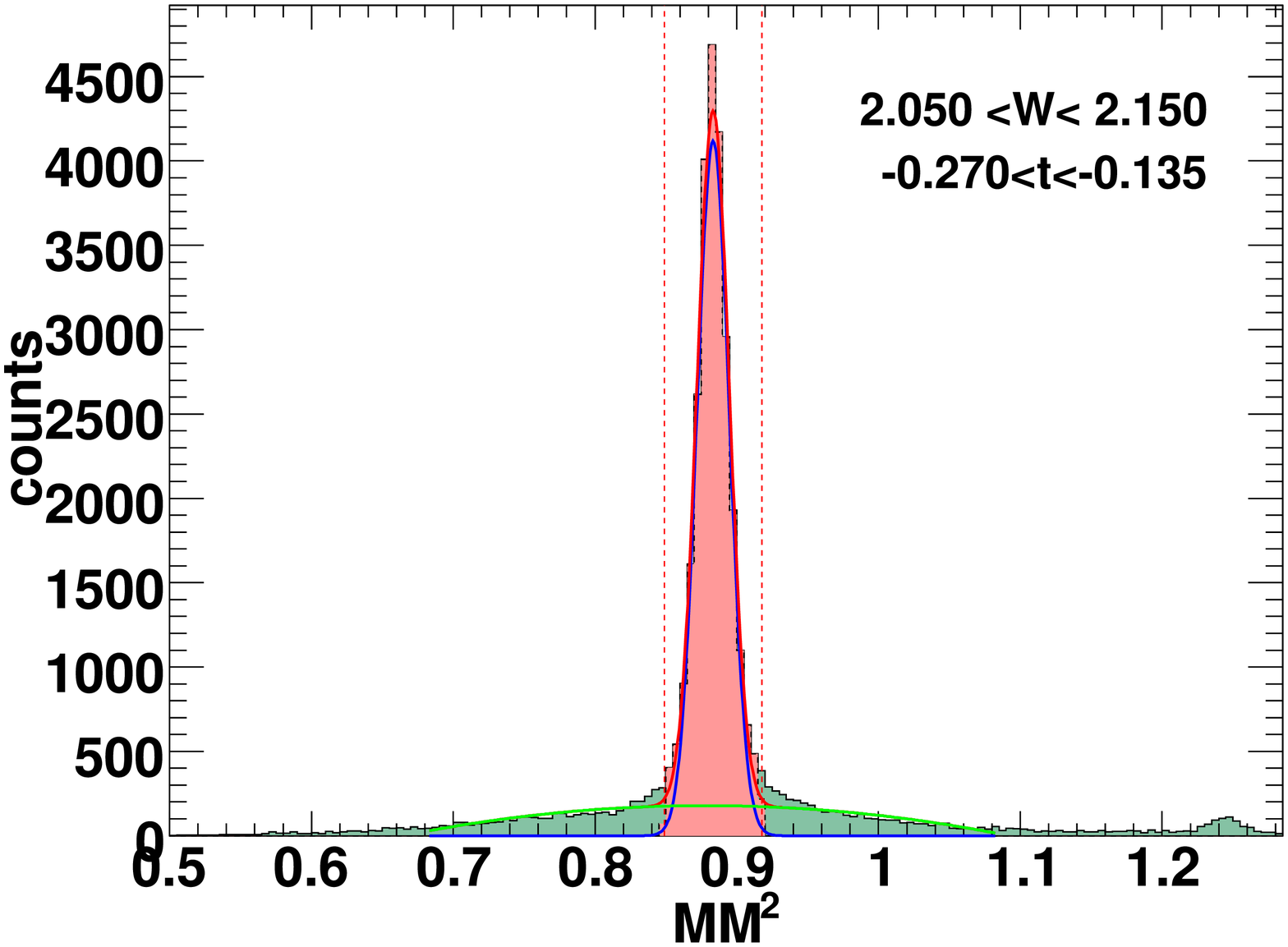}
    \label{fig:case3MM2}
  }
  \subfigure[]{
    \includegraphics[width=.40\textwidth]{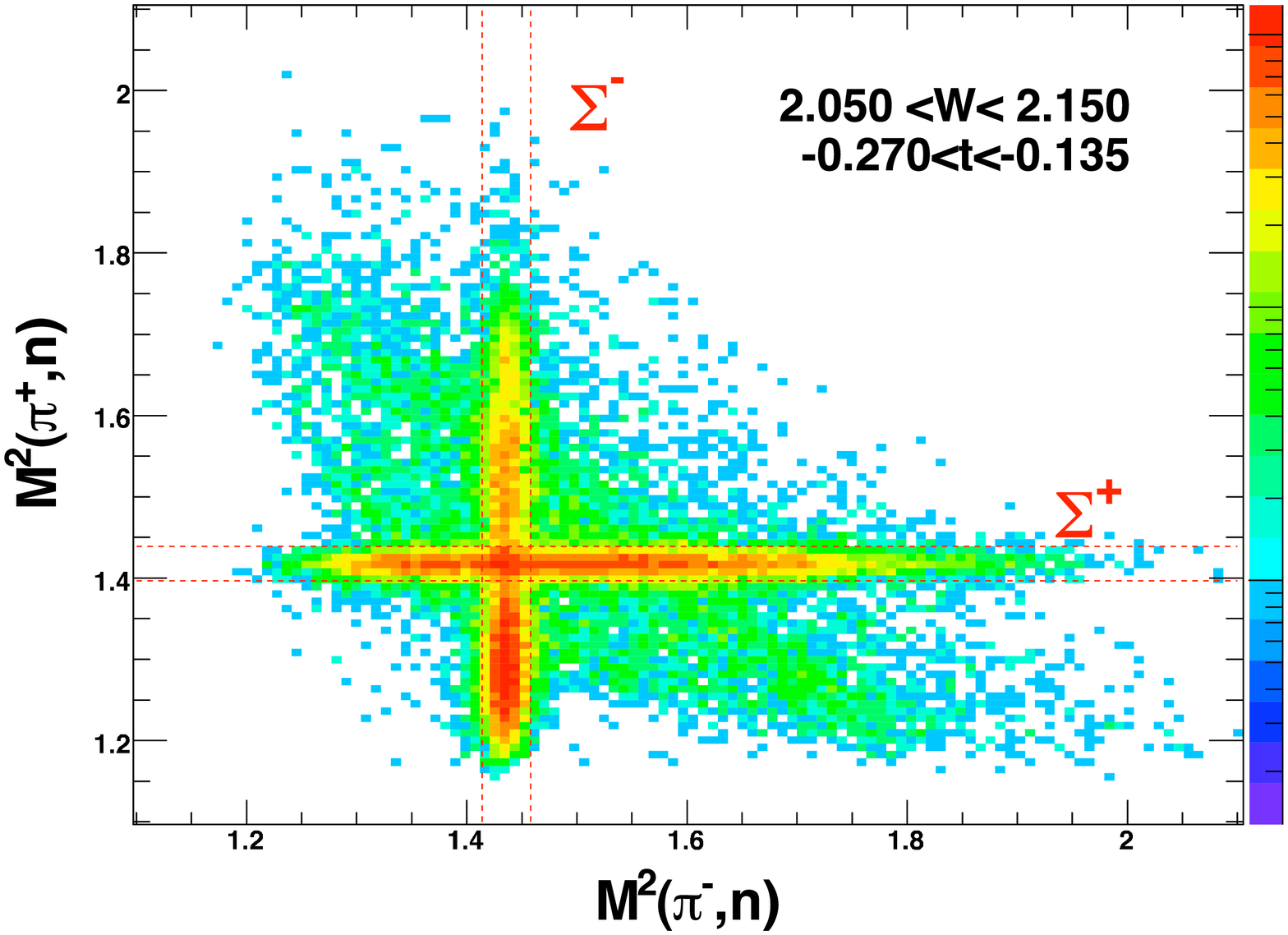}
    \label{fig:case3cross}
  }
\label{fig:analysisplots}
\caption{(Color online)
  \subref{fig:case3MM2} The missing mass squared (\MMsq) calculated
  for one energy and 
  angle bin for the reaction $\gamma + \proton \rightarrow \kp + \pip
  + \pim + X$, where $X$ denotes the missing particle(s). The
  neutron peak is clearly seen. The fit result of a Gaussian (red
  line) and a polynomial background (green line) are also
  shown. The $\pm 3 \sigma$ range of the fit is shown with the red
  dashed lines.
  \subref{fig:case3cross} $M^{2}(\mathrm{n},\pip)$ vs
  $M^{2}(\mathrm{n},\pim)$ for a particular bin
  when detecting $\kp,\pip,\pim$ and reconstructing the missing
  neutron. The horizontal band corresponds to \SigmaPlus{} events,
  and the vertical band to \SigmaMinus{} events. A faint diagonal
  band of neutral kaons decaying to \pip,\pim{} is also seen. $\pm 2
  \sigma$ selection cuts are shown for each hyperon.}
\end{figure}

After the three charged particles are selected and identified using
momentum and TOF information, the
missing mass of the event is calculated to obtain the missing
particle(s). Figure~\ref{fig:case3MM2} shows an example of the missing
mass squared (\MMsq) for one bin of our analysis. A Gaussian fit to
the neutron peak had a $\pm 3 \sigma$ cut applied.
Next, we combine our
final state particles to reconstruct the intermediate ground state
hyperon that the reaction went through. In Figure~\ref{fig:case3cross}
we show the result of plotting the invariant masses squared of the
(n, \pip) versus (n, \pim). Again a fit was done and a region of $\pm
2 \sigma$ was selected around each peak; the $\Sigma$ overlap region
was excluded as well.


At this stage we are ready to fit our resulting $\Sigma \pi$ spectrum as a
sum of templates of Monte Carlo (MC) spectra. The MC were initially
generated with mass and width values given in the PDG~\cite{PDG} and
processed through the standard CLAS simulation package to account for
detection ineffeciencies.
After this fitting process, the non-\LambdaOne{} contributions were
subtracted off from the data and the ``residual'' spectrum was taken
to be from the $\Lambda(1405)$. The resulting lineshape was then acceptance
corrected bin by bin in mass. The final lineshape spectrum was then
obtained by summing over all the production angle bins in
each energy bin. Our final lineshapes are thus shown this way,
acceptance corrected and summed over all production angles.

Once the above process of fitting and obtaining the lineshapes in each
energy bin was completed, the resulting lineshapes were then
used as input to generate MC templates for the \LambdaOne{}, so that
the process could be iterated and refined. The fit result in one
sample bin is shown in Figure \ref{fig:lineshapefit} for the iterated
version.


\section{Results}
\label{sec:results}

\begin{figure}[htp]
  \centering
  \subfigure[]{
    \includegraphics[width=.48\textwidth]{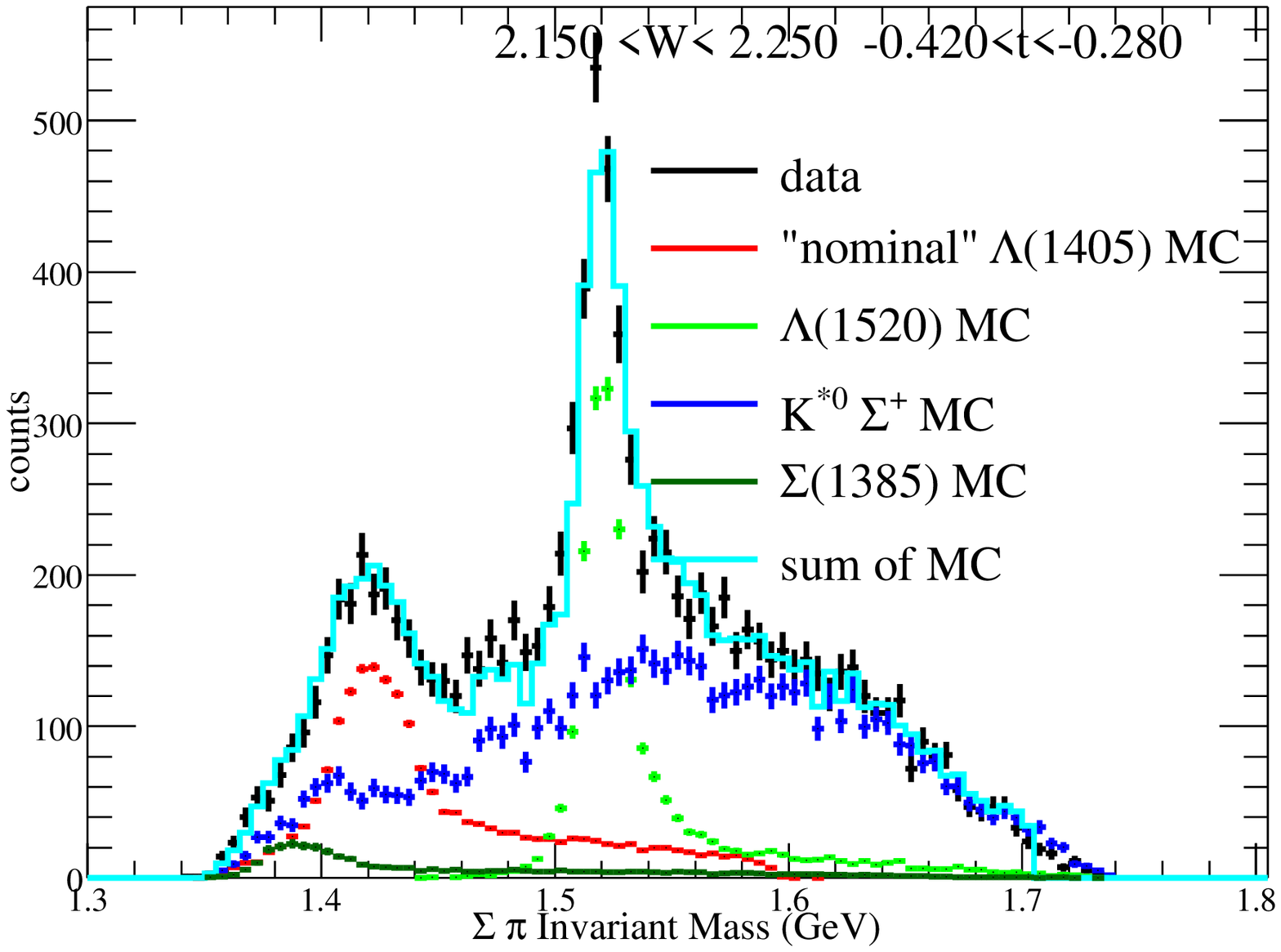}
    \label{fig:lineshapefit}
  }
  \subfigure[]{
    \includegraphics[width=.48\textwidth]{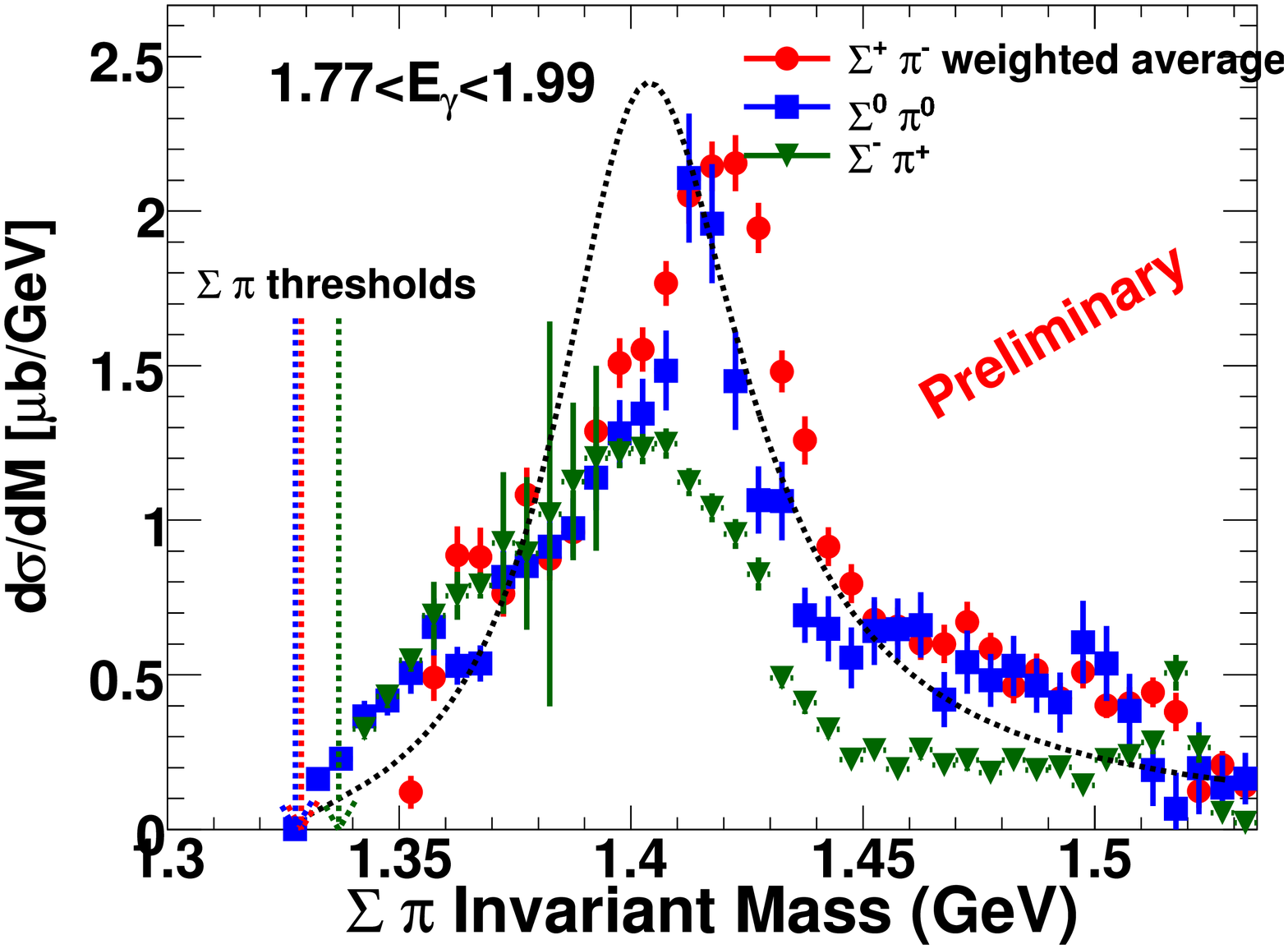}
    \label{fig:lineshape}
  }
  \label{fig:fitandresult}
  \caption{(Color online)
    \subref{fig:lineshapefit}
    Sample of fit to $M(\Sigma \pi)$ spectrum in one bin of the
    $\SigmaPlus \rightarrow \mathrm{n} \pip$ decay mode. Only the
    heights of the
    MC templates were allowed to change. The $\Sigma(1385)$ yield was
    measured in the $\Lambda \pizero$ decay mode, and scaled by
    branching ratio into the $\Sigma \pi$ decay modes. This figure is
    for the iterated version of the fit, see text for details.
    \subref{fig:lineshape}
    The final \LambdaOne{} lineshape result that we obtained for
      one energy bin near threshold. The three different lineshapes are
      plotted against each other. Stark differences are observed,
      especially between the two charged decay modes. The dashed line is
      a relativistic Breit-Wigner shape using the nominal values of mass
      and width of the \LambdaOne{} given in the PDG. Notice that
      although the \SigmaPlus \pim{} threshold is lower than that of
      \SigmaMinus \pip, the \SigmaMinus \pip{} spectrum appears at a lower
      mass.
  }
\end{figure}

After the iteration of the fit, the mass distribution we
obtained is shown in Figure~\ref{fig:lineshape} for one energy bin
near threshold of the \LambdaOne.
Our lineshape result is shown for the three different charge combinations
\SigmaPlus \pim (red), \SigmaZero \pizero (blue), and \SigmaMinus
\pip 
(green). The \SigmaPlus \pim{} spectrum is a weighted average of the two
different decay channels of the \SigmaPlus. The difference in the
lineshapes is clear, with the \SigmaPlus \pim{} spectrum peaking at the
highest mass, and having a much sharper structure than the \SigmaMinus
\pip{} spectrum.



A comparison of our experimental result can be done with the
theoretical prediction (Figure~\ref{fig:theorylineshape}). Although
the lineshapes differ from each other as predicted, we see that for
the theory
prediction, the \SigmaMinus \pip{} spectrum is shifted toward the higher
mass, whereas this applies to the \SigmaPlus \pim{} spectrum in our
experimental result. The difference in the two charged decay modes,
according to the theory, is based on an interference between isospin
$I=0$ and $I=1$ terms, as shown in \cite{Nacher}. At the moment, this
``interchange'' of lineshapes is an unresolved issue that we hope to
resolve in future studies.

\section{Future Prospects and Conclusion}
\label{sec:future-prospects}

The mass distributions of the \LambdaOne{} have been obtained in nine different
energy bins, and although much work has been put into obtaining these,
there is still room for improvement. One example of prospective work
is using the kinematic fitter that was developed for CLAS. The results
shown here do not utilize the kinematic fitter, but our preliminary
studies show that our resolution is readily improved when using it.

Preliminary differential cross sections for the \LambdaOne{} were also
shown during the presentation, which were obtained by summing over
each lineshape in each energy and angle bin. The values of this
\dsigmadt showed differences between the different charge combinations, not
just in scale but also in dependence on $t$. The nature of this
different $t$-dependence for each decay mode is unknown.

Another prospect for our analysis is direct experimental evidence for
the spin and parity of the $\Lambda(1405)$. Although the majority of
theories assume $J^{P} = \frac{1}{2}^{-}$, no direct experimental
proof for these quantum numbers has been given~\cite{spinparity}. Our
preliminary analysis shows a detectable polarization for the
$\Lambda(1405)$, which is transferred to the \SigmaPlus, allowing us to
infer its spin and parity. Our results are consistent with the
common expectation, but will nontheless be an important step
in identifying the nature of the $\Lambda(1405)$. Also, it should be noted
that from these measurements the polarization of the \LambdaOne{} is
measured, giving further information about its production mechanism.

In conclusion, an overview of the CLAS analysis of the \LambdaOne{}
has been 
given. We have an unprecedented high statistics data set for the
$\Lambda(1405)$, and have obtained the lineshapes and differential cross
sections in energy bins spanning from threshold up to $2.84$ GeV in
the overall center of mass energy. We
observe a difference in the different decay mode lineshapes, leading
to a difference in the differential cross sections. We have just
started comparing our results to theory, and are eager to finalize our
results in the coming months.



\end{document}